\documentclass[twocolumn]{aastex6}
\usepackage{color}
\usepackage{savesym}
\savesymbol{tablenum}
\usepackage{amsmath,amssymb,siunitx,ulem,hyperref,multirow}
\restoresymbol{aug}{tablenum}

\newcommand{\prx}{Phys. Rev. X}

\shorttitle{On the possibility of GW190425 being a black hole--neutron
star binary merger}
\shortauthors{Kyutoku et al.}

\begin{document}

\title{On the possibility of GW190425 being a black hole--neutron star
binary merger}
\author{
Koutarou Kyutoku\altaffilmark{1,2,3,4},
Sho Fujibayashi\altaffilmark{5},
Kota Hayashi\altaffilmark{2},
Kyohei Kawaguchi\altaffilmark{6},
Kenta Kiuchi\altaffilmark{5,2},
Masaru Shibata\altaffilmark{5,2},
Masaomi Tanaka\altaffilmark{7}
}
\altaffiltext{1}{Department of Physics, Kyoto University, Kyoto
606-8502, Japan}
\altaffiltext{2}{Center for Gravitational Physics, Yukawa Institute for
Theoretical Physics, Kyoto University, Kyoto 606-8502, Japan}
\altaffiltext{3}{Theory Center, Institute of Particles and Nuclear
Studies, KEK, Tsukuba, Ibaraki 305-0801, Japan}
\altaffiltext{4}{Interdisciplinary Theoretical and Mathematical Sciences
Program (iTHEMS), RIKEN, Wako, Saitama 351-0198, Japan}
\altaffiltext{5}{Max Planck Institute for Gravitational Physics (Albert
Einstein Institute), Am M{\"u}hlenberg 1, Potsdam-Golm D-14476, Germany}
\altaffiltext{6}{Institute for Cosmic Ray Research, The University of
Tokyo, 5-1-5 Kashiwanoha, Kashiwa, Chiba 277-8582, Japan}
\altaffiltext{7}{Astronomical Institute, Tohoku University, Aoba, Sendai
980-8578, Japan}

\begin{abstract}
 We argue that the kilonova/macronova associated with the
 gravitational-wave event GW190425 could have been bright enough to be
 detected if it was caused by the merger of a low-mass black hole and a
 neutron star. Although tidal disruption occurs for such a low-mass
 black hole as is generally expected, the masses of the dynamical ejecta
 are limited to $\lesssim\num{e-3}M_\odot$, which is consistent with
 previous work in the literature. The remnant disk could be as massive
 as $0.05$--$0.1M_\odot$, and the disk outflow of
 $\sim0.01$--$0.03M_\odot$ is likely to be driven by viscous or
 magnetohydrodynamic effects. The disk outflow may not be neutron-rich
 enough to synthesize an abundance of lanthanide elements, even in the
 absence of strong neutrino emitter, if the ejection is driven on the
 viscous time scale of $\gtrsim\SI{0.3}{\second}$. If this is the case,
 the opacity of the disk outflow is kept moderate, and a
 kilonova/macronova at the distance of GW190425 reaches a detectable
 brightness of $20$--$\SI{21}{mag}$ at \SI{1}{day} after merger for most
 viewing angles. If some disk activity ejects the mass within
 $\sim\SI{0.1}{\second}$, instead, lanthanide-rich outflows would be
 launched and the detection of emission becomes challenging. Future
 possible detections of kilonovae/macronovae from GW190425-like systems
 will disfavor the prompt collapse of binary neutron stars and a
 non-disruptive low-mass black hole--neutron star binary associated with
 a small neutron-star radius, whose mass ejection is negligible. The
 host-galaxy distance will constrain the viewing angle and deliver
 further information about the mass ejection.
\end{abstract}

\maketitle

\section{Introduction} \label{sec:intro}

Gravitational-wave observations continue to refine our understanding of
compact objects. The first event and subsequent detections of
binary-black-hole mergers revealed the ubiquity of massive black holes
\citep{ligovirgo2019-3}, some of which may be spinning rapidly
\citep{venumadhav_zrdz}. The first detection of a binary-neutron-star
merger, GW170817 \citep{ligovirgo2017-3} followed by extensive
electromagnetic observations
\citep{ligovirgoem2017,ligovirgogamma2017,goldstein_etal2017,savchenko_etal2017,arcavi_etal2017,coulter_etal2017,lipunov_etal2017,soaressantos_etal2017,tanvir_etal2017,valenti_syctcjrhk2017}
delivered a wealth of information about supranuclear-density matter
\citep[e.g.,][]{de_flbbb2018,ligovirgo2018,narikawa_ukkkst}, gamma-ray
bursts
\citep[e.g.,][]{ligovirgogamma2017,goldstein_etal2017,mooley_etal2018},
\textit{r}-process elements
\citep[e.g.,][]{kasen_mbqr2017,tanaka_etal2017}, the theory of
gravitation \citep[e.g.,][]{ligovirgogamma2017,ligovirgo2019-4}, and
cosmological expansion \citep[e.g.,][]{ligovirgoem2017-2}, while their
masses and spins are consistent with the Galactic population
\citep{farrow_zhu_thrane2019}.

The recent report of gravitational waves from a neutron-star binary
system with $\sim3.4M_\odot$, GW190425 (formerly S190425z), may change
our understanding of the mass spectrum of compact objects
\citep{ligovirgo2020}. As the total mass of Galactic binary neutron
stars has been limited to $2.5$--$2.9M_\odot$
\citep{farrow_zhu_thrane2019}, this event represents a population that
has never been found in the Galaxy as noted in \citet{ligovirgo2020}, if
it is genuinely a binary-neutron-star merger. The possible mass
asymmetry could add further value to this event, because neutron stars
as massive as $\gtrsim1.9M_\odot$ have been found only in neutron
star--white dwarf binaries \citep[e.g.,][]{cromartie_etal2019}. On
another front, the violation of the low-spin prior, with which
dimensionless spin parameter is limited to $\le0.05$
\citep{ligovirgo2020}, is not an issue for a black hole. Indeed, if the
lighter component is a neutron star with a typical mass of
$1.35M_\odot$, the heavy one is consistent with $\gtrsim2M_\odot$ and is
not safely concluded to be a neutron star. If this heavy component is a
low-mass black hole, it challenges existence of the mass gap between
black holes and neutron stars
\citep{kreidberg_bfk2012,ozel_pnv2012}. Thus, the determination of the
binary type is essential to obtain robust astrophysical knowledge
\citep[see][for related work in this direction]{barbieri_scgpc2019}.

It is generally difficult to differentiate low-mass\footnote{In this
Letter, ``low mass'' means $\gtrsim1.9M_\odot$ for which the distinction
from the neutron star is not trivial.} black hole--neutron star binaries
from binary neutron stars. In fact, even GW170817 and associated
electromagnetic counterparts have not been strictly confirmed to be a
binary-neutron-star merger
\citep{foucart_dknps2019,hinderer_etal2019,coughlin_dietrich2019}. The
distinction of the binary type becomes more difficult in the absence of
electromagnetic counterparts.

Unfortunately, no plausible electromagnetic counterpart was detected
with GW190425. This is because of either intrinsically dim emission or
insufficient coverage of the huge error region (see Table 2 of
\citet{hosseinzadeh_etal2019} for the compilation of relevant follow-up
observations). Despite the poor localization of GW190425, however, the
Global Relay of Observatories Watching Transients Happen network
observed impressive thousands of square degrees, which amount to 46\%
and 21\% of the 90\% probability region derived by BAYESTER and
LALInference, respectively \citep{coughlin_etal2019}. Intriguingly, it
is claimed in \citet{pozanenko_mgc2019} that GRB 190425 coincident with
GW190425 had come from the northern hemisphere covered by
\citet{coughlin_etal2019}. It would be worthwhile to investigate whether
such an optical survey has sufficient discriminative power for future
reference (see also \citet{andreoni_etal,coughlin_dabfhrhn2020} for
related work on GW190425 as a binary-neutron-star merger and other
gravitational-wave candidates).

In this Letter, we discuss whether GW190425 could have been identified
unambiguously as a black hole--neutron star binary or binary neutron
stars, if follow-up observations like \citet{coughlin_etal2019} would
have targeted the right sky location. We base our arguments on a suite
of latest numerical-relativity simulations \citep[Hayashi et~al. in
preparation,][]{fujibayashi_swkks2020} for the mass ejection and
radiation-transfer simulations for the kilonova/macronova
\citep{kawaguchi_st}. Note that none of the simulations are tuned to
reproduce this specific event, GW190425.

\section{Gravitational waves} \label{sec:gw}

First, we recall that gravitational-wave observations with the current
sensitivity of detectors are unlikely to differentiate black
hole--neutron star binaries from binary neutron stars, whereas it is
possible and concrete in principle.

\subsection{Inspiral phase}

It is broadly recognized that gravitational waves from the inspiral
phase are not useful to distinguish low-mass black hole--neutron star
binaries from binary neutron stars with current detectors
\citep[e.g.,][]{hannam_bffh2013}. The primary reason is the $\sim99.9\%$
correlation between the mass ratio and the spin parameter
\citep{cutler_flanagan1994}. A precise determination of these
degenerated parameters and an associated classification of binary types
will become possible only when gravitational-wave detectors improve
their sensitivity or multiband gravitational-wave astronomy is realized
\citep{sesana2016}.

Even if the degeneracy is solved, it is intrinsically difficult to
determine whether an object with $\gtrsim2M_\odot$ is a black hole or a
neutron star from our limited knowledge about the neutron-star maximum
mass \citep{shibata_zkf2019}. Although tidal deformability and the
spin-induced quadrupole parameter of a neutron star are different from
those of a black hole, their extraction is challenging
\citep{harry_hinderer2018}. This is particularly the case for a neutron
star near the maximum mass, where these values approach the black-hole
limit \citep{yagi_yunes2013}.

\subsection{Merger phase}

Another possible method of distinction is to observe gravitational waves
from the merger phase in detail. If the source is a low-mass black
hole--neutron star binary, tidal disruption of the neutron star suddenly
shuts off gravitational-wave emission without exciting quasi-normal
modes \citep{shibata_kyt2009}. If the source is massive binary neutron
stars with a total mass of $\sim3.4M_\odot$, the plausible merger
outcome is the prompt collapse \citep{hotokezaka_kosk2011}, and the
ringdown emission from the remnant black hole should accompany it. Our
simulations indicate that this is also true for a highly asymmetric
binary for which the light component is tidally disrupted
\citep{kiuchi_kst2019}. The difference in waveforms is evident for the
case when a massive neutron star is formed.

However, the merger phase can be detected only with high sensitivity at
high frequency. This is because the characteristic frequency of both
tidal disruption and quasi-normal modes are in the kilohertz range
\citep[e.g.,][]{shibata_kyt2009}. Thus, gravitational waves are not
likely to enable us to differentiate a low-mass black hole--neutron star
binary and binary neutron stars until the third-generation detectors
come online.

\section{Electromagnetic counterpart} \label{sec:em}

Next, we discuss whether electromagnetic counterparts allow us to
distinguish a low-mass black hole--neutron star binary and binary
neutron stars for a GW190425-like event. Specifically, we focus on the
observation of the kilonova/macronova down to $\sim\SI{21}{mag}$ at
1--\SI{2}{day} after merger in the \textit{r}- and \textit{g}-bands
typically achieved in \citet{coughlin_etal2019}. They also put an upper
limit of $\sim\SI{15}{mag}$ in the \textit{J}-band, but this is not
restrictive in realistic situations.

\subsection{Mass ejection}

The properties of the kilonova/macronova are determined by the ejected
material \citep{li_paczynski1998}. Thus, we begin with investigating the
mass ejection from GW190425-like systems based on our
numerical-relativity simulations. The ejection channels are usually
separated into (hydro)dynamical processes working during merger and disk
activity in the post-merger phase, and accordingly we discuss these two
separately below.

First of all, we recall that no mass ejection is expected from
symmetric\footnote{In this Letter, ``symmetric'' means the mass ratio of
$0.9$--$1$, where the precise threshold is admittedly ambiguous.}
binary neutron stars as massive as $\sim3.4M_\odot$, because they are
highly likely to collapse promptly into a black hole without leaving
baryonic material
\citep{kiuchi_sst2010,hotokezaka_kkosst2013}. Accordingly, no
kilonova/macronova is expected for this case as well as gamma-ray
bursts. This means that any non-detection will be consistent with
massive binary neutron stars as far as the (nearly-)equal-mass system is
allowed by gravitational-wave observations. If a massive neutron star is
formed transiently helped by a large maximum mass, the black hole--disk
formation and mass ejection could be similar to those for asymmetric
binaries \citep{kiuchi_kst2019}.\footnote{This scenario strongly
indicates that the merger remnant of GW170817 survived for long period
of time.} Hereafter in this section, we focus on clarifying the
similarity and difference between a low-mass black hole--neutron star
binary and asymmetric binary neutron stars.

\subsubsection{Merger phase}

Coincidentally, we have recently simulated mergers of non-spinning
$2M_\odot$ black hole--$1.35M_\odot$ neutron star binaries, which
reasonably model possible sources of GW190425 (Hayashi et~al. in
preparation). Data analysis of GW190425 indicates that the spin
parameter is small \citep{ligovirgo2020}, and the neglect of the spin is
acceptable. We adopted three piecewise polytropes, HB, H, and 1.25H
\citep{lackey_ksbf2012},\footnote{1.25H is denoted by $p6.0\Gamma3.0$ in
\citet{lackey_ksbf2012}.} with which neutron-star radii took
\SI{11.6}{\km}, \SI{12.3}{\km}, and \SI{13.0}{\km}, respectively. This
range spans middle to large radii inferred by GW170817
\citep{ligovirgo2018} and should be sufficient to cover non-trivial
outcomes of black hole--neutron star binary mergers. Actual simulations
are performed with the SACRA-MPI code
\citep{yamamoto_st2008,kiuchi_kksst2017} with three different
resolutions to check numerical convergence. Because these merger
simulations aim at deriving the amounts of dynamical ejecta and formed
accretion disks, neutrino transport or magnetic fields are not
incorporated (see below for post-merger simulations).

The masses of the dynamical ejecta are found to be as small as
$<\num{e-4}M_\odot$, $<\num{e-4}M_\odot$, and $\num{4e-4}M_\odot$ for
HB, H, and 1.25H, respectively, although the neutron stars are disrupted
violently to leave moderately massive disks of $0.04M_\odot$,
$0.07M_\odot$, and $0.1M_\odot$. The smallness of the dynamical ejecta
in low-mass black hole--neutron star binaries is consistent with the
tendency found in our previous study (see Fig.~11 of
\citet{kyutoku_iost2015} and also \citet{foucart_dknps2019}). Although
we did not incorporate neutrino transport in these simulations, previous
studies confirmed that the dynamical ejecta are extremely neutron-rich
for black hole--neutron star binaries
\citep{roberts_ldfflnop2017,kyutoku_ksst2018}.

We have also simulated $1.8M_\odot$--$1.2M_\odot$ mergers of black
hole--neutron star binaries and binary neutron stars with the APR4
equation of state \citep{akmal_pr1998}, with which the radius of a
$1.2M_\odot$ neutron star is \SI{11.0}{\km}, for comparing outcomes
(Hayashi et~al. in preparation). We find that (i) the mass of the disk,
$0.07$--$0.08M_\odot$, depends only weakly on the binary type, and (ii)
the mass of the dynamical ejecta is $\num{9e-4}M_\odot$ for asymmetric
binary neutron stars, which is larger by an order of magnitude than that
for corresponding black hole--neutron star binaries. The masses of the
dynamical ejecta owe their difference to the presence/absence of the
surface of the heavier neutron star, which allows the material of the
disrupted lighter component to be ejected before the whole system
collapses into the remnant black hole \citep[see
also][]{kiuchi_kst2019}. While these models are not directly comparable
to GW190425, our results suggest that the dynamical ejecta of
$\gtrsim\num{e-3}M_\odot$ was possible if GW190425 was asymmetric binary
neutron stars, particularly if the neutron-star radius is
$\gtrsim\SI{12}{\km}$.

To summarize, the dynamical ejecta are very tiny for low-mass black
hole--neutron star binaries, while they could be
$\gtrsim\num{e-3}M_\odot$ for asymmetric and massive binary neutron
stars. By contrast, the disk mass can be as large as
$\sim0.05$--$0.1M_\odot$ irrespective of the binary type unless the
neutron-star radius is $\lesssim\SI{11.5}{\km}$. This indicates that the
main ejection channel for these systems is likely to be the post-merger
disk activity.

\subsubsection{Post-merger phase}

It is commonly believed that the outflow from the black hole--accretion
disk is neutron-rich enough to synthesize abundance of lanthanide
elements
\citep[e.g.,][]{just_bagj2015,siegel_metzger2018,fernandez_tqfk2019}. This
should be contrasted with the lanthanide-poor outflow from the tori
surrounding massive neutron stars, which emit a copious amount of
neutrinos to mitigate neutron richness
\citep{fujibayashi_knss2018}. However, no simulation of the black
hole--disk outflow has been performed in general-relativistic viscous
hydrodynamics. In addition, initial conditions have typically been given
without referring to merger simulations \citep[but see
also][]{fernandez_fkldr2017}.

To explore properties of the outflow from the remnant black
hole--accretion disk, we performed axisymmetric
viscous-neutrino-radiation-hydrodynamics simulations in full general
relativity \citep{fujibayashi_swkks2020}. Our fiducial models consist of
a $0.1M_\odot$ disk surrounding a $3M_\odot$ black hole with its
dimensionless spin parameter being $0.8$ or $0.6$. This reasonably
models the merger product of GW190425 for the case of a low-mass black
hole--neutron star binary or asymmetric binary neutron stars (see the
previous section for the disk mass).

The simulations show that 15\%--30\% of the initial disk material is
ejected on the time scale of $\sim0.3$--$\SI{1}{\second}$ in a
quasi-spherical manner for a plausible range of the alpha viscosity
parameter, $\alpha=0.05$--$0.15$. This ejection efficiency agrees with
that found in previous work
\citep[e.g.,][]{just_bagj2015,siegel_metzger2018,fernandez_tqfk2019}. The
outflow is driven by the viscosity, which is presumably induced by
magneto-turbulence in realistic situations, and the contribution of
neutrino-driven winds is negligible. Because the disk outflow dominates
the dynamical ejecta in terms of the mass, properties of the disk
outflow primarily determine whether the kilonova/macronova is detectable
for a GW190425-like system, irrespective of the binary type.

One crucial difference of our results from previous ones
\citep[e.g.,][]{just_bagj2015,siegel_metzger2018,fernandez_tqfk2019} is
that the electron fraction $Y_\mathrm{e}$ of the disk outflow is not
very low with its mass-averaged value being $\sim0.3$ even in the
absence of the strong neutrino emitter, i.e., the remnant neutron
star. This result is rather similar to that of our previous work on
accretion tori around massive neutron stars
\citep{fujibayashi_knss2018}. The reason for this difference is that the
mass ejection occurs on the viscous time scale of
$\sim0.3$--\SI{1}{\second} as stated above. Because the outflow is not
driven significantly until the weak processes freeze out and neutrino
cooling becomes inefficient, the disk material typically relaxes to
equilibrium of electron/positron captures by the time of ejection. Thus,
the neutron richness is mitigated for most of the disk outflow
\citep{fujibayashi_swkks2020}. By contrast, if the mass ejection occurs
on a short time scale of $\lesssim\SI{0.1}{\second}$, the neutron
richness of the ejecta could be high
\citep[e.g.,][]{siegel_metzger2018,fernandez_tqfk2019,christie_ltffqk2019}.

Post-process nuclear network calculations show that the lanthanide mass
fraction, a key parameter to determine the opacity
\citep{kasen_mbqr2017,tanaka_kgk}, of the disk outflow derived in our
simulations is limited to $\lesssim\num{e-3}$ for
$(0<)\alpha\lesssim0.1$ \citep{fujibayashi_swkks2020}. The opacity of
such lanthanide-poor ejecta is lower by an order of magnitude than the
lanthanide-rich case. As a general trend, the lanthanide mass fraction
increases as the alpha viscosity parameter increases or the initial mass
of the disk decreases. This trend implies that the lanthanide mass
fraction can have diversity for a given mass of the outflow, and further
case-by-case studies or first-principle
neutrino-radiation-magnetohydrodynamics simulations are warranted.

To summarize, the outflow from the black hole--accretion disk can be
either lanthanide-poor or rich reflecting the time scale of mass
ejection. In the following, we consider both cases for assessing the
detectability of electromagnetic counterparts. Meanwhile, the ejection
efficiency is likely to be 15\%--30\%.

We note that the similarity of the remnant black hole--disk system
suggests that possible gamma-ray bursts should also be similar,
irrespective of the binary type. However, it is premature to conclude
anything about the driving process of an ultrarelativistic jet. We do
not discuss the gamma-ray burst here, and it definitely requires future
investigation.

\subsection{Kilonova/Marconova}

\begin{figure*}
 \plotone{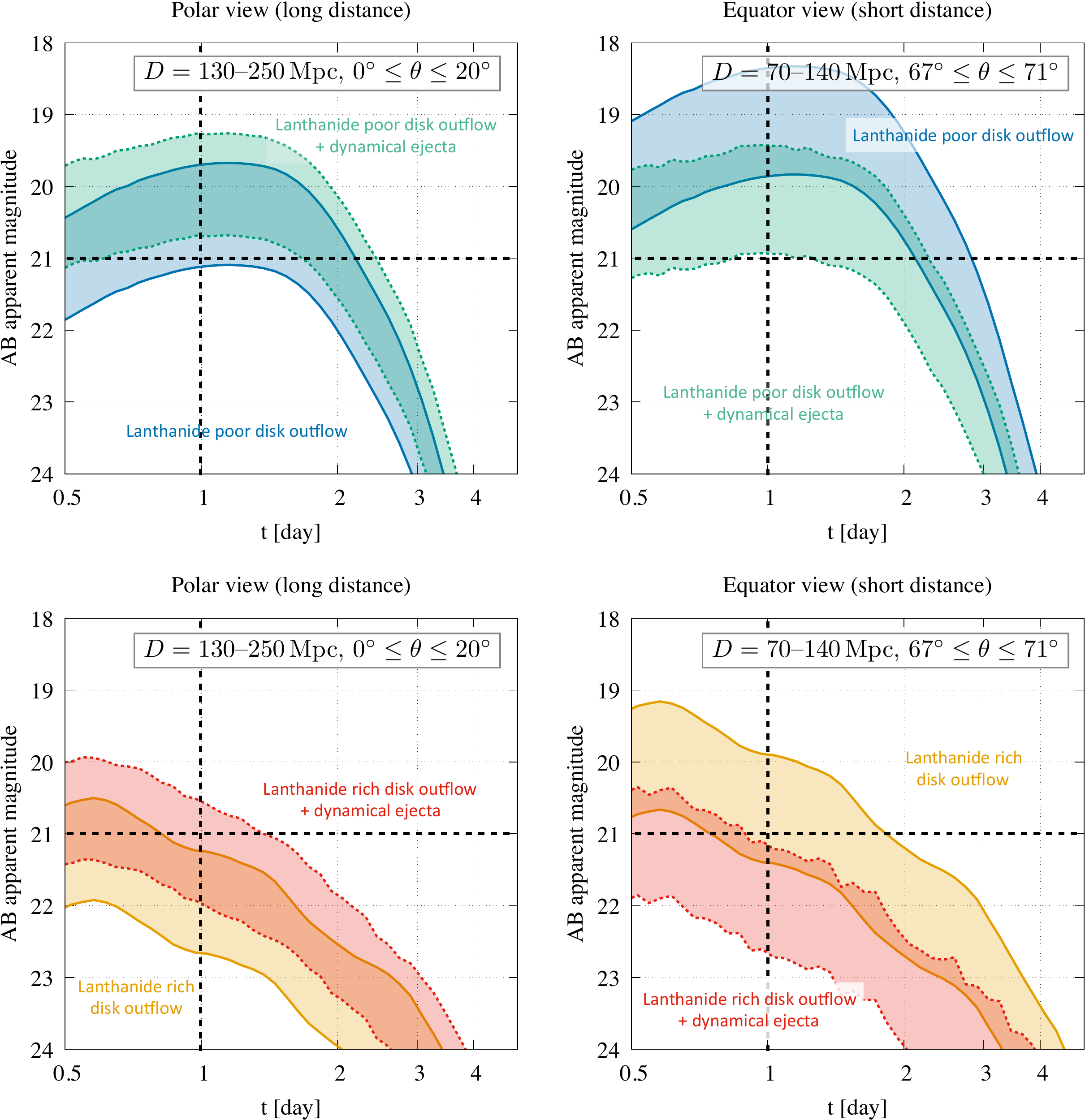} \caption{Various possible $r$-band light curves in AB
 magnitude of the kilonova/macronova for GW190425-like events. The left
 and right columns show results for polar viewing angles of
 \ang{0}--\ang{20} from the rotational axis of the binary at
 $130$--\SI{250}{Mpc} and the equatorial one of \ang{67}--\ang{71} at
 $70$--\SI{140}{Mpc}, respectively. The distance is matched to GW190425
 taking the correlation with the viewing angle into account
 \citep{ligovirgo2020}. The blue and green strips in the top panels
 correspond to the lanthanide-poor disk outflows of $0.02M_\odot$
 without and with dynamical ejecta of $\num{e-4}M_\odot$,
 respectively. The orange and red strips in the bottom panels are the
 same as above but for the lanthanide-rich disk outflows. Thick dashed
 lines indicate \SI{21}{mag} and \SI{1}{day} \citep{coughlin_etal2019}.}
 \label{fig:lc}
\end{figure*}

To assess the detectability, we computed multiband light curves of the
kilonova/macronova associated with various possible configurations of
ejecta from GW190425-like systems by Monte-Carlo radiation-transfer
simulations with the code developed in
\citet{tanaka_hotokezaka2013,kawaguchi_st}, and the line list described
in \citet{tanaka_kgk}. The density profile and anisotropy of the
dynamical ejecta are modeled following \citet{kawaguchi_st}. The
mass-averaged velocity of the disk outflow is varied from 6\% to 10\% of
the speed of light, and the following discussion does not depend on the
specific choice. It should be cautioned that predictions of the
kilonova/macronova depend on computational methods and systematic errors
should be mitigated by cross comparisons with other models \citep[see
e.g.,][]{kasen_mbqr2017,bulla2019,coughlin_dabfhrhn2020}.

The results are summarized in Fig.~\ref{fig:lc} for two representative
viewing angles with incorporating distance uncertainties, which depend
on the viewing angle via the correlation in determining
gravitational-wave amplitude \citep{ligovirgo2020}. Specifically, the
distance should be larger for more polar direction, and vice versa, for
given amplitude. We explain individual cases below.

Our simulations indicate that a survey equivalent to
\citet{coughlin_etal2019} is likely to detect the emission from the disk
outflow if it is lanthanide poor (see the top row of
Fig.~\ref{fig:lc}). Specifically, for the material with
$Y_\mathrm{e}=0.3$--$0.4$, the emission at \SI{1}{day} after merger
reaches $\sim\SI{20}{mag}$ in the \textit{r/g}-bands for a hypothetical
distance of $\SI{150}{Mpc}$. Indeed, the emission reaches $\SI{21}{mag}$
even if the mass is reduced to $0.01M_\odot$. Thus, if the non-detection
of \citet{coughlin_etal2019} had really covered the location of
GW190425, neither a low-mass black hole--neutron star binary nor
asymmetric binary neutron stars is preferred unless a small neutron-star
radius of $\lesssim\SI{11.5}{\km}$ suppresses the disk
formation. Although the dynamical ejecta with $\gtrsim\num{e-4}M_\odot$
could conceal the emission in the equatorial direction by the
lanthanide-curtain effect \citep{kasen_fm2015} even taking the proximity
into account, this applies only to highly equatorial angles of
$\gtrsim\ang{75}$ and half of the azimuthal angle due to anisotropy
\citep{kyutoku_iost2015}.

If the disk outflow is lanthanide rich, the detectability depends on the
combination of the viewing angle and the amount of the dynamical ejecta
(see the bottom row of Fig.~\ref{fig:lc}). Specifically, for the
material with $Y_\mathrm{e}=0.1$--$0.3$, the high opacity renders the
emission from the outflow of $0.02M_\odot$ marginally undetectable
$\SI{21.5}{mag}$ for $\SI{150}{Mpc}$. This can be detected from the
equatorial direction with a small distance, but not from the polar
direction with a large distance. However, the lanthanide-rich dynamical
ejecta of $\gtrsim\num{e-4}M_\odot$ associated with a large neutron-star
radius of $\gtrsim\SI{12.5}{\km}$ interchange the detectability of these
two cases. On the one hand, the emission in the polar direction is
boosted by $0.5$--$\SI{1}{mag}$ and could become detectable. On the
other hand, the lanthanide-rich disk outflow is concealed by the
lanthanide-rich dynamical ejecta and becomes undetectable in the
equatorial direction despite their similar composition. This is ascribed
to the low temperature and density of the dynamical ejecta, where the
latter increases the expansion opacity as far as the medium is optically
thick \citep[see e.g., Equation (1) of][]{kawaguchi_st}. Because
gravitational-wave detections are biased toward polar directions
\citep{schutz2011}, the dynamical ejecta may tend to enhance the
detectability.

We caution that enhancement in the polar direction due to the dynamical
ejecta is only marginal for the cases relevant to GW190425, particularly
taking uncertain microphysics in radiation-transfer simulations into
account \citep{kawaguchi_st}. For the \textit{r}-band magnitude to
exceed $\SI{21}{mag}$ significantly for $\SI{150}{Mpc}$, the dynamical
ejecta as massive as $\gtrsim0.01M_\odot$ are required, which is
comparable to the mass of the disk outflow. Such vigorous dynamical mass
ejection is unlikely to occur for low-mass black hole--neutron star
binaries. This might be possible with asymmetric binary neutron stars,
whose kilonova/macronova may be characterized by the disk outflow with
relatively massive dynamical ejecta.

\section{Discussion} \label{sec:discussion}

\begin{deluxetable*}{ccc}
 \tablewidth{0pt}
 \tablecaption{Detectability of the kilonovae/macronovae from GW190425-like
 events with respect to the binary type and the merger outcome
 \label{tab:score}}
 \tablehead{\colhead{Binary type\tablenotemark{a}} & \colhead{Merger
 Outcome\tablenotemark{b}} & \colhead{Detectable?\tablenotemark{c}}}
 \startdata
 \multirow{5}{*}{low-mass BH--NS} & La-poor disk & YES \\
  & La-poor disk$+$La-rich dyn. & $\approx$YES \\
  & La-rich disk & YES if equatorial \\
  & La-rich disk$+$La-rich dyn. & YES if polar \\
  & weak/no disruption (small radius) & NO \\ \hline
  \multirow{2}{*}{asymmetric NS--NS} & La-poor disk$+$La-rich dyn. & YES
 if polar \\
  & La-rich disk$+$La-rich dyn. & YES if polar \\ \hline
  \multirow{2}{*}{symmetric NS--NS} & massive neutron star (large
 maximum mass) & YES if polar \\
  & prompt collapse & NO \\
 \enddata
 \tablecomments{Parentheses in the ``Merger outcome'' column indicate
 the required neutron-star properties. Any binary type becomes trivially
 undetectable if the ejection efficiency happens to be low.}
 \tablenotetext{a}{BH: black hole; NS: neutron star}
 \tablenotetext{b}{La-rich: lanthanide rich; La-poor: lanthanide poor;
 disk: disk outflow of $\sim\num{e-2}M_\odot$; dyn.: dynamical ejecta of
 $\gtrsim\num{e-4}M_\odot$}
 \tablenotetext{c}{$\approx$YES: except for highly equatorial directions
 heavily obscured by lanthanide curtains; polar: observed from the polar
 direction avoiding lanthanide curtains; equatorial: observed from the
 equatorial direction with a small distance}
 \end{deluxetable*}

Although huge uncertainties are inevitable, we speculate implications of
possible future detections of kilonovae/macronovae from GW190425-like
events by a survey like \citet{coughlin_etal2019} (see Table
\ref{tab:score} for the summary of detectability). Because a successful
detection will tell us the host galaxy and its distance, accuracy in
determining the viewing angle will be improved by mitigating the
degeneracy in the gravitational-wave amplitude \citep[see e.g.,][for the
case of GW170817]{mandel2018,finstad_dbbb2018}. First, if it turns out
that we are in the equatorial direction of the event, the dynamical
ejecta may be $\lesssim\num{e-4}M_\odot$ for the disk outflow to avoid
lanthanide curtains. This case may support the hypothesis of a low-mass
black hole--neutron star binary if massive and luminous dynamical ejecta
of $\gtrsim0.01M_\odot$ are rejected by further follow-up
observations. If the dynamical ejecta are massive, asymmetric binary
neutron stars are favored. Second, if we are in the polar direction, the
lanthanide-rich disk outflow without dynamical ejecta is not
preferred. This may disfavor lanthanide-rich disk outflows after tidal
disruption of neutron stars with small radii of $\lesssim\SI{12.5}{\km}$
by low-mass black holes. Any detection will not support binary neutron
stars resulting in prompt collapse or a low-mass black hole--neutron
star binary with a small neutron-star radius of
$\lesssim\SI{11.5}{\km}$, because they leave a negligible amount of
material.

If an upper limit of $\sim\SI{21}{mag}$ is established for a future
GW190425-like event, the most likely source may be symmetric binary
neutron stars that collapsed promptly, and a low-mass black
hole--neutron star binary may be acceptable only for the following cases
(see Table \ref{tab:score} for the summary): (i) it is observed from the
equatorial direction, thus close, but the dynamical ejecta associated
with a large neutron-star radius of $\gtrsim\SI{12.5}{\km}$ concealed
the lanthanide-rich disk outflow, and the same applies to the
lanthanide-poor disk outflow for highly equatorial directions; (ii) it
is observed from the polar direction, thus distant, and early ejection
of the lanthanide-rich disk outflow occurs without significant dynamical
mass ejection due to a small neutron-star radius of
$\lesssim\SI{12.5}{\km}$, or (iii) the disk outflow is suppressed to
$\lesssim0.01M_\odot$ due to insignificant disk formation associated
with a small neutron-star radius of $\lesssim\SI{11.5}{\km}$ \citep[see
also, e.g.,][for preference of a large radius by pulsar
observations]{miller_etal2019} or accidentally low ejection efficiency
from the disk.

If GW190425-like systems are not very rare, we have a good chance of
observing them with the multiple detector network in the near future
\citep{lvk2018}. The localization error should be improved to the extent
that follow-up observations will be able to cover the entire
localization area, hopefully deeper. For a hypothetical distance of
$\SI{150}{Mpc}$ consistent with GW190425, the peak magnitude of the
kilonova/macronova may reach $\approx\SI{20}{mag}$ in the
\textit{r/g}-bands for asymmetric systems, powered mainly by the disk
outflow. Thus, observations comparable to \cite{coughlin_etal2019} will
detect the emission or put an informative limit on the possibility of
low-mass black hole--neutron star binaries (and asymmetric binary
neutron stars). The multiple detector network also improves the accuracy
in the distance and the viewing angle \citep{cutler_flanagan1994}. Even
if the localization error is not improved due to, e.g., low duty cycle,
future sensitive surveys with telescopes like the Vera~C. Rubin
Observatory (formerly LSST) will be a powerful tool to detect
GW190425-like events. To differentiate binary types with future
multi-messenger astronomy, it will be beneficial to improve
understanding of the neutron-star equation of state and disk activity.
Studies on these topics will also improve phenomenological models of the
kilonova/macronova employed in expansive statistical analysis
\citep[e.g.,][]{coughlin_dietrich2019,coughlin_dmm2019,hinderer_etal2019,coughlin_dabfhrhn2020},
which will play an important role for quantitative inferences of source
properties.

\acknowledgments
Numerical computations were performed at Oakforest-PACS at Information
Technology Center of the University of Tokyo, Cray XC50 at CfCA of
National Astronomical Observatory of Japan, Cray XC40 at Yukawa
Institute for Theoretical Physics of Kyoto University, and Sakura and
Cobra clusters at Max Planck Computing and Data Facility. This work is
supported by Japanese Society for the Promotion of Science (JSPS)
KAKENHI grant No.~JP16H02183, No.~JP16H06342, No.~JP17H01131,
No.~JP18H01213, No.~JP18H04595, No.~JP18H05236, and No.~JP19K14720.


\begin{thebibliography}{}
\expandafter\ifx\csname natexlab\endcsname\relax\def\natexlab#1{#1}\fi

\bibitem[{{Abbott} {et~al.}(2017{\natexlab{a}}){Abbott}, {Abbott}, {Abbott},
  {Acernese}, {Ackley}, {Adams}, {Adams}, {Addesso}, {Adhikari}, {Adya}, \&
  et~al.}]{ligovirgoem2017-2}
{Abbott}, B.~P., {Abbott}, R., {Abbott}, T.~D., {et~al.} 2017{\natexlab{a}},
  \nat, 551, 85

\bibitem[{{Abbott} {et~al.}(2017{\natexlab{b}}){Abbott}, {Abbott}, {Abbott},
  {Acernese}, {Ackley}, {Adams}, {Adams}, {Addesso}, {Adhikari}, {Adya}, \&
  et~al.}]{ligovirgogamma2017}
---. 2017{\natexlab{b}}, \apjl, 848, L13

\bibitem[{{Abbott} {et~al.}(2017{\natexlab{c}}){Abbott}, {Abbott}, {Abbott},
  {Acernese}, {Ackley}, {Adams}, {Adams}, {Addesso}, {Adhikari}, {Adya}, \&
  et~al.}]{ligovirgo2017-3}
---. 2017{\natexlab{c}}, \prl, 119, 161101

\bibitem[{{Abbott} {et~al.}(2017{\natexlab{d}}){Abbott}, {Abbott}, {Abbott},
  {Acernese}, {Ackley}, {Adams}, {Adams}, {Addesso}, {Adhikari}, {Adya}, \&
  et~al.}]{ligovirgoem2017}
---. 2017{\natexlab{d}}, \apjl, 848, L12

\bibitem[{{Abbott} {et~al.}(2018{\natexlab{a}}){Abbott}, {Abbott}, {Abbott},
  {Acernese}, {Ackley}, {Adams}, {Adams}, {Addesso}, {Adhikari}, {Adya}, \&
  et~al.}]{ligovirgo2018}
---. 2018{\natexlab{a}}, \prl, 121, 161101

\bibitem[{{Abbott} {et~al.}(2018{\natexlab{b}}){Abbott}, {Abbott}, {Abbott},
  {Abernathy}, {Acernese}, {Ackley}, {Adams}, {Adams}, {Addesso}, {Adhikari},
  \& et~al.}]{lvk2018}
---. 2018{\natexlab{b}}, Living Reviews in Relativity, 21, 3

\bibitem[{{Abbott} {et~al.}(2019{\natexlab{a}}){Abbott}, {Abbott}, {Abbott},
  {Acernese}, {Ackley}, {Adams}, {Adams}, {Addesso}, {Adhikari}, {Adya}, \&
  et~al.}]{ligovirgo2019-3}
---. 2019{\natexlab{a}}, \prx, 9, 031040

\bibitem[{{Abbott} {et~al.}(2019{\natexlab{b}}){Abbott}, {Abbott}, {Abbott},
  {Acernese}, {Ackley}, {Adams}, {Adams}, {Addesso}, {Adhikari}, {Adya}, \&
  et~al.}]{ligovirgo2019-4}
---. 2019{\natexlab{b}}, \prl, 123, 011102

\bibitem[{{Abbott} {et~al.}(2020){Abbott}, {Abbott}, {Abbott}, {Abraham},
  {Acernese}, {Ackley}, {Adams}, {Adhikari}, {Adya}, {Affeldt}, \&
  et~al.}]{ligovirgo2020}
---. 2020, arXiv:2001.01761

\bibitem[{Akmal {et~al.}(1998)Akmal, Pandharipande, \&
  Ravenhall}]{akmal_pr1998}
Akmal, A., Pandharipande, V.~R., \& Ravenhall, D.~G. 1998, \prc, 58, 1804

\bibitem[{{Andreoni} {et~al.}(2019){Andreoni}, {Goldstein}, {Kasliwal},
  {Nugent}, {Zhou}, {Newman}, {Bulla}, {Foucart}, {Hotokezaka}, {Nakar}, \&
  et~al.}]{andreoni_etal}
{Andreoni}, I., {Goldstein}, D.~A., {Kasliwal}, M.~M., {et~al.} 2019,
  arXiv:1910.13409

\bibitem[{{Arcavi} {et~al.}(2017){Arcavi}, {Hosseinzadeh}, {Howell}, {McCully},
  {Poznanski}, {Kasen}, {Barnes}, {Zaltzman}, {Vasylyev}, {Maoz}, \&
  et~al.}]{arcavi_etal2017}
{Arcavi}, I., {Hosseinzadeh}, G., {Howell}, D.~A., {et~al.} 2017, \nat, 551, 64

\bibitem[{Barbieri {et~al.}(2019)Barbieri, Salafia, Colpi, Ghirlanda, Perego,
  \& Colombo}]{barbieri_scgpc2019}
Barbieri, C., Salafia, O.~S., Colpi, M., {et~al.} 2019, \apjl, 887, L35

\bibitem[{Bulla(2019)}]{bulla2019}
Bulla, M. 2019, \mnras, 489, 5037

\bibitem[{Christie {et~al.}(2019)Christie, Lalakos, Tchekhovskoy,
  Fern{\'a}ndez, Foucart, Quataert, \& Kasen}]{christie_ltffqk2019}
Christie, I.~M., Lalakos, A., Tchekhovskoy, A., {et~al.} 2019, \mnras, 490,
  4811

\bibitem[{Coughlin \& Dietrich(2019)}]{coughlin_dietrich2019}
Coughlin, M.~W., \& Dietrich, T. 2019, \prd, 100, 043011

\bibitem[{Coughlin {et~al.}(2019)Coughlin, Dietrich, Margalit, \&
  Metzger}]{coughlin_dmm2019}
Coughlin, M.~W., Dietrich, T., Margalit, B., \& Metzger, B.~D. 2019, \mnras,
  489, L91

\bibitem[{{Coughlin} {et~al.}(2019){Coughlin}, {Ahumada}, {Anand}, {De},
  {Hankins}, {Kasliwal}, {Singer}, {Bellm}, {Andreoni}, {Cenko}, \&
  et~al.}]{coughlin_etal2019}
{Coughlin}, M.~W., {Ahumada}, T., {Anand}, S., {et~al.} 2019, \apjl, 885, L19

\bibitem[{Coughlin {et~al.}(2020)Coughlin, Dietrich, Antier, Bulla, Foucart,
  Hotokezaka, Raaijmakers, Hinderer, \& Nissanke}]{coughlin_dabfhrhn2020}
Coughlin, M.~W., Dietrich, T., Antier, S., {et~al.} 2020, \mnras, 492, 863

\bibitem[{{Coulter} {et~al.}(2017){Coulter}, {Foley}, {Kilpatrick}, {Drout},
  {Piro}, {Shappee}, {Siebert}, {Simon}, {Ulloa}, {Kasen}, \&
  et~al.}]{coulter_etal2017}
{Coulter}, D.~A., {Foley}, R.~J., {Kilpatrick}, C.~D., {et~al.} 2017, Science,
  358, 1556

\bibitem[{{Cromartie} {et~al.}(2019){Cromartie}, {Fonseca}, {Ransom},
  {Demorest}, {Arzoumanian}, {Blumer}, {Brook}, {DeCesar}, {Dolch}, {Ellis}, \&
  et~al.}]{cromartie_etal2019}
{Cromartie}, H.~T., {Fonseca}, E., {Ransom}, S.~M., {et~al.} 2019, Nature
  Astronomy, 4, 72

\bibitem[{Cutler \& Flanagan(1994)}]{cutler_flanagan1994}
Cutler, C., \& Flanagan, {\'E}.~E. 1994, \prd, 49, 2658

\bibitem[{De {et~al.}(2018)De, Finstad, Lattimer, Brown, Berger, \&
  Biwer}]{de_flbbb2018}
De, S., Finstad, D., Lattimer, J.~M., {et~al.} 2018, \prl, 121, 091102

\bibitem[{Farrow {et~al.}(2019)Farrow, Zhu, \& Thrane}]{farrow_zhu_thrane2019}
Farrow, N., Zhu, X.-J., \& Thrane, E. 2019, \apj, 876, 18

\bibitem[{Fern{\'a}ndez {et~al.}(2017)Fern{\'a}ndez, Foucart, Kasen, Lippuner,
  Desai, \& Roberts}]{fernandez_fkldr2017}
Fern{\'a}ndez, R., Foucart, F., Kasen, D., {et~al.} 2017, Classical and Quantum
  Gravity, 34, 154001

\bibitem[{Fern{\'a}ndez {et~al.}(2019)Fern{\'a}ndez, Tchekhovskoy, Quataert,
  Foucart, \& Kasen}]{fernandez_tqfk2019}
Fern{\'a}ndez, R., Tchekhovskoy, A., Quataert, E., Foucart, F., \& Kasen, D.
  2019, \mnras, 482, 3373

\bibitem[{Finstad {et~al.}(2018)Finstad, De, Brown, Berger, \&
  Biwer}]{finstad_dbbb2018}
Finstad, D., De, S., Brown, D.~A., Berger, E., \& Biwer, C.~M. 2018, \apjl,
  860, L2

\bibitem[{Foucart {et~al.}(2019)Foucart, Duez, Kidder, Nissanke, Pfeiffer, \&
  Scheel}]{foucart_dknps2019}
Foucart, F., Duez, M.~D., Kidder, L.~E., {et~al.} 2019, \prd, 99, 103025

\bibitem[{Fujibayashi {et~al.}(2018)Fujibayashi, Kiuchi, Nishimura, Sekiguchi,
  \& Shibata}]{fujibayashi_knss2018}
Fujibayashi, S., Kiuchi, K., Nishimura, N., Sekiguchi, Y., \& Shibata, M. 2018,
  \apj, 860, 64

\bibitem[{Fujibayashi {et~al.}(2020)Fujibayashi, Shibata, Wanajo, Kiuchi,
  Kyutoku, \& Sekiguchi}]{fujibayashi_swkks2020}
Fujibayashi, S., Shibata, M., Wanajo, S., {et~al.} 2020, arXiv:2001.04467

\bibitem[{{Goldstein} {et~al.}(2017){Goldstein}, {Veres}, {Burns}, {Briggs},
  {Hamburg}, {Kocevski}, {Wilson-Hodge}, {Preece}, {Poolakkil}, {Roberts}, \&
  et~al.}]{goldstein_etal2017}
{Goldstein}, A., {Veres}, P., {Burns}, E., {et~al.} 2017, \apjl, 848, L14

\bibitem[{Hannam {et~al.}(2013)Hannam, Brown, Fairhurst, Fryer, \&
  Harry}]{hannam_bffh2013}
Hannam, M., Brown, D.~A., Fairhurst, S., Fryer, C.~L., \& Harry, I.~W. 2013,
  \apjl, 766, L14

\bibitem[{Harry \& Hinderer(2018)}]{harry_hinderer2018}
Harry, I., \& Hinderer, T. 2018, Classical and Quantum Gravity, 35, 145010

\bibitem[{Hinderer {et~al.}(2019)Hinderer, Nissanke, Foucart, Hotokezaka,
  Vincent, Kasliwal, Schmidt, Williamson, Nichols, Duez, Kidder, Pfeiffer, \&
  Scheel}]{hinderer_etal2019}
Hinderer, T., Nissanke, S., Foucart, F., {et~al.} 2019, \prd, 100, 063021

\bibitem[{{Hosseinzadeh} {et~al.}(2019){Hosseinzadeh}, {Cowperthwaite},
  {Gomez}, {Villar}, {Nicholl}, {Margutti}, {Berger}, {Chornock}, {Paterson},
  {Fong}, \& et~al.}]{hosseinzadeh_etal2019}
{Hosseinzadeh}, G., {Cowperthwaite}, P.~S., {Gomez}, S., {et~al.} 2019, \apjl,
  880, L4

\bibitem[{Hotokezaka {et~al.}(2013)Hotokezaka, Kiuchi, Kyutoku, Okawa,
  Sekiguchi, Shibata, \& Taniguchi}]{hotokezaka_kkosst2013}
Hotokezaka, K., Kiuchi, K., Kyutoku, K., {et~al.} 2013, \prd, 87, 024001

\bibitem[{Hotokezaka {et~al.}(2011)Hotokezaka, Kyutoku, Okawa, Shibata, \&
  Kiuchi}]{hotokezaka_kosk2011}
Hotokezaka, K., Kyutoku, K., Okawa, H., Shibata, M., \& Kiuchi, K. 2011, \prd,
  83, 124008

\bibitem[{Just {et~al.}(2015)Just, Bauswein, Pulpillo, Goriely, \&
  Janka}]{just_bagj2015}
Just, O., Bauswein, A., Pulpillo, R.~A., Goriely, S., \& Janka, H.-T. 2015,
  \mnras, 448, 541

\bibitem[{Kasen {et~al.}(2015)Kasen, Fern{\'a}ndez, \& Metzger}]{kasen_fm2015}
Kasen, D., Fern{\'a}ndez, R., \& Metzger, B.~D. 2015, \mnras, 450, 1777

\bibitem[{Kasen {et~al.}(2017)Kasen, Metzger, Barnes, Quataert, \&
  Ramirez-Ruiz}]{kasen_mbqr2017}
Kasen, D., Metzger, B., Barnes, J., Quataert, E., \& Ramirez-Ruiz, E. 2017,
  \nat, 551, 80

\bibitem[{Kawaguchi {et~al.}(2019)Kawaguchi, Shibata, \& Tanaka}]{kawaguchi_st}
Kawaguchi, K., Shibata, M., \& Tanaka, M. 2019, arXiv:1908.05815

\bibitem[{Kiuchi {et~al.}(2017)Kiuchi, Kawaguchi, Kyutoku, Sekiguchi, Shibata,
  \& Taniguchi}]{kiuchi_kksst2017}
Kiuchi, K., Kawaguchi, K., Kyutoku, K., {et~al.} 2017, \prd, 96, 084060

\bibitem[{Kiuchi {et~al.}(2019)Kiuchi, Kyutoku, Shibata, \&
  Taniguchi}]{kiuchi_kst2019}
Kiuchi, K., Kyutoku, K., Shibata, M., \& Taniguchi, K. 2019, \apjl, 876, L31

\bibitem[{Kiuchi {et~al.}(2010)Kiuchi, Sekiguchi, Shibata, \&
  Taniguchi}]{kiuchi_sst2010}
Kiuchi, K., Sekiguchi, Y., Shibata, M., \& Taniguchi, K. 2010, \prl, 104,
  141101

\bibitem[{Kreidberg {et~al.}(2012)Kreidberg, Bailyn, Farr, \&
  Kalogera}]{kreidberg_bfk2012}
Kreidberg, L., Bailyn, C.~D., Farr, W.~M., \& Kalogera, V. 2012, \apj, 757, 36

\bibitem[{Kyutoku {et~al.}(2015)Kyutoku, Ioka, Okawa, Shibata, \&
  Taniguchi}]{kyutoku_iost2015}
Kyutoku, K., Ioka, K., Okawa, H., Shibata, M., \& Taniguchi, K. 2015, \prd, 92,
  044028

\bibitem[{Kyutoku {et~al.}(2018)Kyutoku, Kiuchi, Sekiguchi, Shibata, \&
  Taniguchi}]{kyutoku_ksst2018}
Kyutoku, K., Kiuchi, K., Sekiguchi, Y., Shibata, M., \& Taniguchi, K. 2018,
  \prd, 97, 023009

\bibitem[{Lackey {et~al.}(2012)Lackey, Kyutoku, Shibata, Brady, \&
  Friedman}]{lackey_ksbf2012}
Lackey, B.~D., Kyutoku, K., Shibata, M., Brady, P.~R., \& Friedman, J.~L. 2012,
  \prd, 85, 044061

\bibitem[{Li \& Paczy{\'n}ski(1998)}]{li_paczynski1998}
Li, L.-X., \& Paczy{\'n}ski, B. 1998, \apjl, 507, L59

\bibitem[{{Lipunov} {et~al.}(2017){Lipunov}, {Gorbovskoy}, {Kornilov},
  {Tyurina}, {Balanutsa}, {Kuznetsov}, {Vlasenko}, {Kuvshinov}, {Gorbunov},
  {Buckley}, \& et~al.}]{lipunov_etal2017}
{Lipunov}, V.~M., {Gorbovskoy}, E., {Kornilov}, V.~G., {et~al.} 2017, \apjl,
  850, L1

\bibitem[{Mandel(2018)}]{mandel2018}
Mandel, I. 2018, \apjl, 853, L12

\bibitem[{{Miller} {et~al.}(2019){Miller}, {Lamb}, {Dittmann}, {Bogdanov},
  {Arzoumanian}, {Gendreau}, {Guillot}, {Harding}, {Ho}, {Lattimer}, \&
  et~al.}]{miller_etal2019}
{Miller}, M.~C., {Lamb}, F.~K., {Dittmann}, A.~J., {et~al.} 2019, \apjl, 887,
  L24

\bibitem[{{Mooley} {et~al.}(2018){Mooley}, {Nakar}, {Hotokezaka}, {Hallinan},
  {Corsi}, {Frail}, {Horesh}, {Murphy}, {Lenc}, {Kaplan}, \&
  et~al.}]{mooley_etal2018}
{Mooley}, K.~P., {Nakar}, E., {Hotokezaka}, K., {et~al.} 2018, \nat, 554, 207

\bibitem[{Narikawa {et~al.}(2019)Narikawa, Uchikata, Kawaguchi, Kiuchi,
  Kyutoku, Shibata, \& Tagoshi}]{narikawa_ukkkst}
Narikawa, T., Uchikata, N., Kawaguchi, K., {et~al.} 2019, arXiv:1910.08971

\bibitem[{{\"O}zel {et~al.}(2012){\"O}zel, Psaltis, Narayan, \&
  Villarreal}]{ozel_pnv2012}
{\"O}zel, F., Psaltis, D., Narayan, R., \& Villarreal, A.~S. 2012, \apj, 757,
  55

\bibitem[{Pozanenko {et~al.}(2019)Pozanenko, Minaev, Grebenev, \&
  Chelovekov}]{pozanenko_mgc2019}
Pozanenko, A.~S., Minaev, P.~Y., Grebenev, S.~A., \& Chelovekov, I.~V. 2019,
  arXiv:1912.13112

\bibitem[{Roberts {et~al.}(2017)Roberts, Lippuner, Duez, Faber, Foucart,
  {Lomberdi, Jr.}, Ning, Ott, \& Ponce}]{roberts_ldfflnop2017}
Roberts, L.~F., Lippuner, J., Duez, M.~D., {et~al.} 2017, \mnras, 464, 3907

\bibitem[{{Savchenko} {et~al.}(2017){Savchenko}, {Ferrigno}, {Kuulkers},
  {Bazzano}, {Bozzo}, {Brandt}, {Chenevez}, {Courvoisier}, {Diehl}, {Domingo},
  \& et~al.}]{savchenko_etal2017}
{Savchenko}, V., {Ferrigno}, C., {Kuulkers}, E., {et~al.} 2017, \apjl, 848, L15

\bibitem[{Schutz(2011)}]{schutz2011}
Schutz, B.~F. 2011, Classical and Quantum Gravity, 28, 125023

\bibitem[{Sesana(2016)}]{sesana2016}
Sesana, A. 2016, \prl, 116, 231102

\bibitem[{Shibata {et~al.}(2009)Shibata, Kyutoku, Yamamoto, \&
  Taniguchi}]{shibata_kyt2009}
Shibata, M., Kyutoku, K., Yamamoto, T., \& Taniguchi, K. 2009, \prd, 79, 044030

\bibitem[{Shibata {et~al.}(2019)Shibata, Zhou, Kiuchi, \&
  Fujibayashi}]{shibata_zkf2019}
Shibata, M., Zhou, E., Kiuchi, K., \& Fujibayashi, S. 2019, \prd, 100, 023015

\bibitem[{Siegel \& Metzger(2018)}]{siegel_metzger2018}
Siegel, D.~M., \& Metzger, B.~D. 2018, \apj, 858, 52

\bibitem[{{Soares-Santos} {et~al.}(2017){Soares-Santos}, {Holz}, {Annis},
  {Chornock}, {Herner}, {Berger}, {Brout}, {Chen}, {Kessler}, {Sako}, \&
  et~al.}]{soaressantos_etal2017}
{Soares-Santos}, M., {Holz}, D.~E., {Annis}, J., {et~al.} 2017, \apjl, 848, L16

\bibitem[{Tanaka \& Hotokezaka(2013)}]{tanaka_hotokezaka2013}
Tanaka, M., \& Hotokezaka, K. 2013, \apj, 775, 113

\bibitem[{Tanaka {et~al.}(2019)Tanaka, Kato, Gaigalas, \&
  Kawaguchi}]{tanaka_kgk}
Tanaka, M., Kato, D., Gaigalas, G., \& Kawaguchi, K. 2019, arXiv:1906.08914

\bibitem[{{Tanaka} {et~al.}(2017){Tanaka}, {Utsumi}, {Mazzali}, {Tominaga},
  {Yoshida}, {Sekiguchi}, {Morokuma}, {Motohara}, {Ohta}, {Kawabata}, {Abe},
  {Aoki}, {Asakura}, {Baar}, {Barway}, {Bond}, {Doi}, {Fujiyoshi}, {Furusawa},
  {Honda}, {Itoh}, {Kawabata}, {Kawai}, {Kim}, {Lee}, {Miyazaki}, {Morihana},
  {Nagashima}, {Nagayama}, {Nakaoka}, {Nakata}, {Ohsawa}, {Ohshima}, {Okita},
  {Saito}, {Sumi}, {Tajitsu}, {Takahashi}, {Takayama}, {Tamura}, {Tanaka},
  {Terai}, {Tristram}, {Yasuda}, \& {Zenko}}]{tanaka_etal2017}
{Tanaka}, M., {Utsumi}, Y., {Mazzali}, P.~A., {et~al.} 2017, \pasj, 69, 102

\bibitem[{{Tanvir} {et~al.}(2017){Tanvir}, {Levan},
  {Gonz{\'a}lez-Fern{\'a}ndez}, {Korobkin}, {Mandel}, {Rosswog}, {Hjorth},
  {D'Avanzo}, {Fruchter}, {Fryer}, \& et~al.}]{tanvir_etal2017}
{Tanvir}, N.~R., {Levan}, A.~J., {Gonz{\'a}lez-Fern{\'a}ndez}, C., {et~al.}
  2017, \apjl, 848, L27

\bibitem[{Valenti {et~al.}(2017)Valenti, Sand, Yang, Cappellaro, Tartaglia,
  Corsi, Jha, Reichart, Haislip, \& Kouprianov}]{valenti_syctcjrhk2017}
Valenti, S., Sand, D.~J., Yang, S., {et~al.} 2017, \apjl, 848, L24

\bibitem[{Venumadhav {et~al.}(2019)Venumadhav, Zackay, Roulet, Dai, \&
  Zaldarriaga}]{venumadhav_zrdz}
Venumadhav, T., Zackay, B., Roulet, J., Dai, L., \& Zaldarriaga, M. 2019,
  arXiv:1904.07214

\bibitem[{Yagi \& Yunes(2013)}]{yagi_yunes2013}
Yagi, K., \& Yunes, N. 2013, \prd, 88, 023009

\bibitem[{Yamamoto {et~al.}(2008)Yamamoto, Shibata, \&
  Taniguchi}]{yamamoto_st2008}
Yamamoto, T., Shibata, M., \& Taniguchi, K. 2008, \prd, 78, 064054

\end{thebibliography}
\end{document}